\documentclass[]{tPHM2e}

\newcommand{\BEQ}{\begin{equation}}
\newcommand{\EEQ}{\end{equation}}
\newcommand{\BEA}{\begin{eqnarray}}
\newcommand{\EEA}{\end{eqnarray}}
\renewcommand{\H}{{\mathcal {H}}} 

\newcommand{\nn}{\nonumber }

\begin{document}
\doi{10.1080/1478643YYxxxxxxxx}
\issn{1478-6443}
\issnp{1478-6435}
\jvol{00} \jnum{00} \jyear{00} \jmonth{}


\articletype{RESEARCH ARTICLE}

\title{The overlap parameter across an inverse first order phase transition in a 3D spin-glass}

\author{M. Paoluzzi$^{\rm a,b}$$^{\ast}$\thanks{$^\ast$Corresponding author. Email: mpaoluzzi@fis.uniroma3.it
\vspace{6pt}}
L. Leuzzi$^{\rm a,c}$
A. Crisanti$^{\rm c}$
\\\vspace{6pt}  $^{\rm a}${\em{
IPCF-CNR, UOS Roma, 
P.le Aldo Moro 2, I-00185 Roma, Italy}
}\\
$^{\rm b}${\em{Dipartimento di Fisica,  Universit\`a di Roma 3,
Via della Vasca Navale 84, I-00146 Roma, Italy}
}\\
$^{\rm c}${\em{Dipartimento di Fisica, Universit\`a "Sapienza",
  P.le Aldo Moro 2, I-00185 Roma, Italy}}
}
\maketitle
\begin{abstract}
We
investigate the thermodynamic phase transition 
taking place in the Blume-Capel
model
in presence of quenched disorder in three dimensions (3D). 
In particular, 
performing Exchange Montecarlo simulations, we study
the behavior of the order parameters 
accross the first order
phase transition and its related coexistence region. This 
transition is an Inverse Freezing.
\end{abstract}
\section{Introduction}
\begin{figure}[!t]
\begin{center}
\includegraphics[width=.5\textwidth]{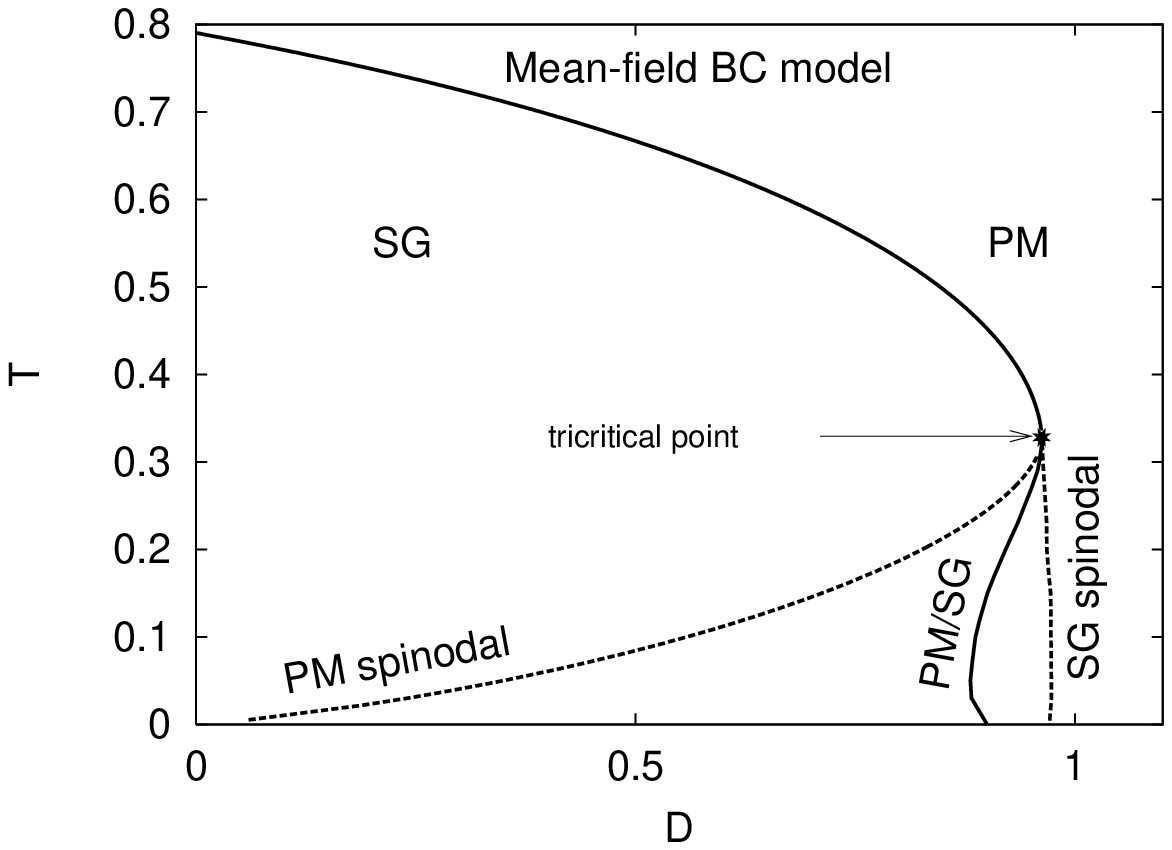}%
\includegraphics[width=.5\textwidth]{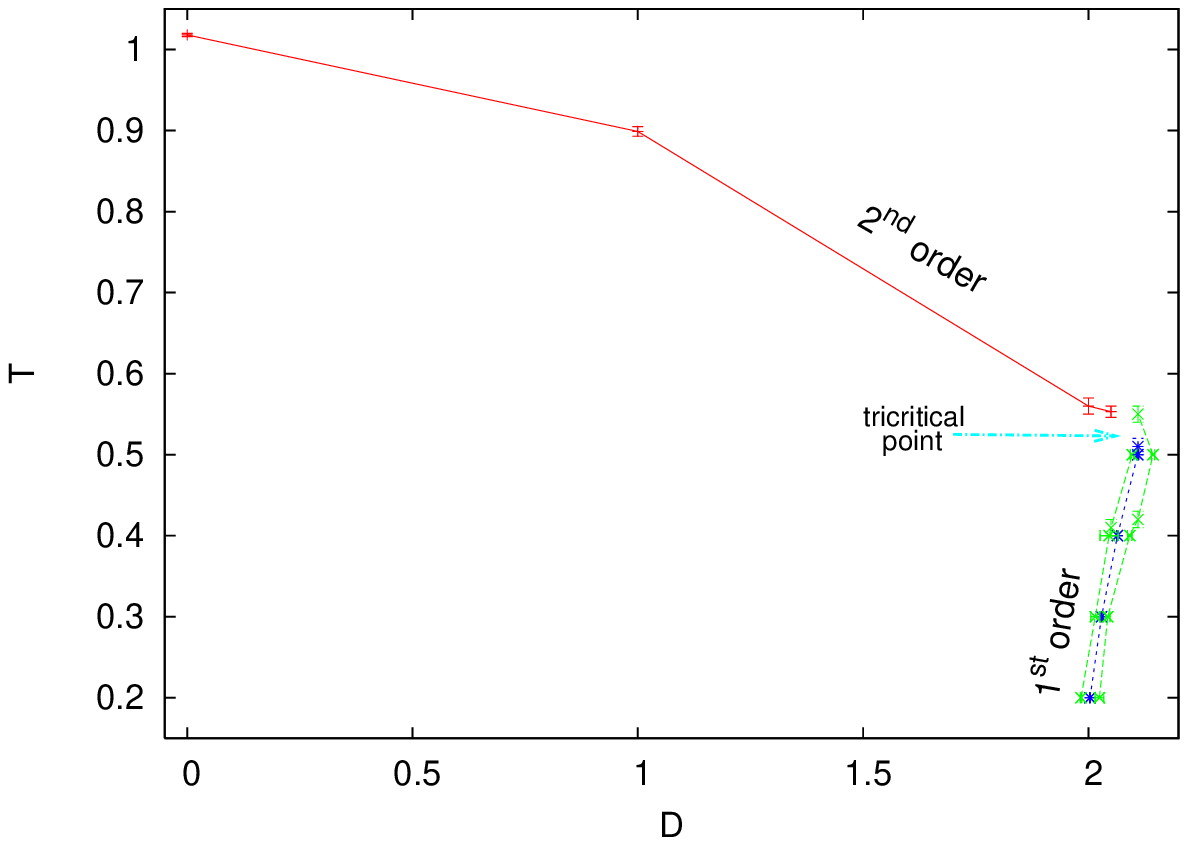}%
\caption{Left panel: the phase diagram of the BC model with
quenched disorder for the fully-connected
lattice with Gaussian distribuited
random couplings (MF solution \cite{CLPRL02,CLPRB04}):
the Second Order Phase Transition ends in a
tricritical point where an Inverse Freezing First
Order Phase Transition takes place.
The variance of $P(J_{ij})$ in the
MF model was $\propto 1/z$, $z$ being
the number of sites connected to each
spin. In the 3D model, where $z=6$ (right panel), a
bimodal distribution has been chosen with
variance $1$ rather than $\propto 1/6$.
}%
\label{mf-pd}
\end{center}
\end{figure}

 \indent In the present paper, we will consider the Blume-Capel\cite{CapelPhys66,BEGPRA71,Berker76} model with
 quenched disorder (BC-random)\cite{GSJPC77,CLPRL02,CLPRB04,Ozcelik08}: 
 a spin-glass model with bosonic
 spin$-1$ variables ($s_i= -1,0,+1$).  
 BC-random is one of the simplest spin glass models that
 displays an {\itshape Inverse Transition} (IT)\cite{Paoluzzi10}. 
 By IT we mean a reversible
transition occurring between phases whose entropic content is in the
inverse order relation relatively to standard transitions. The case -
already hypothesized by Tammann more than a century ago \cite{Tammann}
- of ``ordering in disorder'' taking place in a crystal solid that
liquefies on cooling, is generally termed {\em inverse melting}.
The IT phenomenon also includes the transformation involving amorphous
solid phases, as that of a liquid vitrifying upon heating, and the
term {\em inverse freezing} (IF) is somewhat used in the literature: both
phases are disordered but the fluid appears to be the one with least
entropic content.
IT has been experimentally observed in many materials: 
some examples of inverse melting can be found in\cite{Wilks87,RHKMM99,Greer00,vRRMM04,Plazanet04,Tombari05,Plazanet06,Plazanet06b,Angelini07,Ferrari07,Angelini08,Angelini08b,Plazanet09,Angelini09},
while IF takes place in\cite{CACPS97,Hirrien98,Haque93}.    
The reason for these counter intuitive phenomenon is
that a phase usually at higher entropic content happens to exist in
very peculiar patterns such that its entropy actually decreases below
the entropy of the phase normally considered the most ordered one\cite{Shupper04,Shupper05}.
\\
The Mean Field (MF) solution of the BC-random model in the Full
Replica Symmetry Breaking (RSB) scheme\cite{CLPRL02,CLPRB04} predicts
a phase diagram (fig. \ref{mf-pd}) with a second order transition line (between Spin-Glass (SG) and
Paramagnet (PM) phase) ending in a tricritical point, where a first order phase
transition line starts, and from where a phase coexistence
region departs. 
\\
We stress that the transition is first
order in the {\itshape thermodynamic sense}, with latent heat and is not related
to the so-called {\itshape random first order transition}\cite{RFO} occurring in
MF models for structural glasses. 
Furthermore, the first
order transition is characterized by the phenomenon of IF
\cite{CLPRL05,LPM}: the low temperature phase is PM, with a lower
entropy than the SG phase, and the transition line develops a
reentrance. \\
In the present work we will study the behavior of the 3D BC-random model
on a cubic lattice with nearest-neighbor quenched interaction. 
The nature
of the phases that appear in the phase diagram, in particular across the 
IF First Order Phase Transition (FOPT), is studied through the 
shape of the order
parameters distributions:
this qualitative method allow us to 
understand, in a very simple way, the
fundamental phenomenology
that drives the IF scenario.
Other results on the same model 
have been presented in \cite{Paoluzzi10,Leuzzi10}.
 
\section{Model and Observables}
The Hamiltonian of the BC-random is defined as follows
\BEQ
\H_J[s]=- \sum_{(i,j)}J_{ij} s_i s_j + D\sum_i s_i^2
\EEQ 
where $({i}{j})$ indicates ordered couples of
nearest-neighbor sites, and $s_{{i}}= -1,0,+1$ are spin$-1$
variables lying on a cubic lattice of size $N=L^3$ with Periodic
Boundary Condition. The external crystal field $D$ 
plays the role of a chemical potential. Random couplings
$J_{{i}{j}}$ are independent identically
distributed as
 \BEQ
 P(J_{{i}{j}})=\frac{1}{2}
\delta (J_{{i}{j}}-1)
+\frac{1}{2}\delta(J_{{i}{j}}+1).
 \EEQ
We simulate two real replicas of the system and define the overlap, 
i.e. the order parameter usually characterizing the SG transition, as
 \BEQ
 q^{(J)}\equiv
\frac{1}{N}\sum_{{i}}\langle s^{(1)}_{{i}}
s^{(2)}_{{i}}\rangle
\EEQ
 where $\langle \ldots \rangle$ is the thermal average. If a thermodynamic
first order phase transition occurs, with latent heat, the most significant
  order parameter that drives the transition is the density $\rho$ of
  magnetically active ($|s_i|=1$) sites:
\BEQ
\rho^{(J)}=\frac{1}{N}\sum_{{i}} \langle s^2_{{i}}\rangle\; .
\EEQ 
The apex $J$ recalls us that the values of the parameters depend on
the particular realization of disorder ($\{J_{{i}{j}}\}$). 
All the information about the equilibrium properties of
the system is in the knowledge of the following probability
distribution functions (pdfs)
\BEA
P(q)&\equiv&\overline{P_J(q)} = \overline{\left\langle \delta \left(q^{(J)}-\frac{1}{N}\sum_i s_i^{(1)}s_i^{(2)} \right) \right\rangle} \\ \nn
P(\rho)&\equiv& \overline{P_J(\rho)} = \overline{\left\langle \delta \left(\rho^{(J)}- \frac{1}{N}\sum_{{i}} s^2_{{i}} \right) \right\rangle} \;.
\EEA
We
denote by ${\overline {\phantom{(}\ldots \phantom{)}}}$ the average
over quenched disorder. 
The
dependence on the random couplings $J_{ij}$ is known to be self-averaging for the density probability
distribution, but not for the
overlap distributions  $P_J(q)$, whose average over the quenched
disorder in the thermodynamic limit is different from the
thermodynamic limit of a single realization of random couplings\cite{MPV87}:
\BEA
P(q)&\equiv&\lim_{N\to \infty}
\overline{P_J(q)}\neq \lim_{N\to\infty} P^{(N)}_J(q)\, .
\EEA
We can introduce the notion
of active and inactive site:
when $s_i=0$ the site is inactive; otherwise, if $s_i^2=1$, it 
can interact with its neighbors ($s_i$ is an active site): as we
will see in the next sections the IF indeed takes place 
between a SG of high density to an almost empty PM.
The few active sites practically do not interact with each other 
but almost
exclusively with inactive neighbors 
and this induces zero magnetization and
overlap. The corresponding PM phase at the same $D$ and high $T$ has, instead, higher
density and the paramagnetic behavior is
brought about by the lack of both magnetic order (zero magnetization)
and blocked spin configurations (zero overlap), cf. Sec 3.
\section{Phase Transitions and Order Parameters}
The equilibrium dynamics of the BC-random has been 
numerically simulated through
Parallel Tempering (PT) technique: 
we have simulated in parallel the dynamic of the system at  different values of $T$ and  $D$.
For the PT in $T$, the swap probability of two
copies at $T$ and $T+\Delta T$ was:
\BEQ P_{\mbox{swap}}(\Delta \beta)=
\min\left[1,\exp\{\Delta\beta\Delta\mathcal{H}\}\right]\, . \EEQ
While, for PT in $D$, two copies with $D$ and $D+\Delta D$ were
exchanged with probability 
\BEQ P_{\mbox{swap}}(\Delta D)=\min\left[1,\exp\{\beta\Delta
D\Delta\rho\}\right]\, .\EEQ
We will present data of 3D systems studied with PT in $T$ at $D=0$, 
and in $D$ at $T=0.2,0.3,0.4,0.5$. At $D=0$ we simulated from
$33$ to $40$ replicated copies $N_T$ at linear size
$L=6,8,10,12$ (number of disordered sample: $N_J=2000$), for
$D=0$ we simulated $N_T\in [20:33]$ at $L=16,20$
($N_J\in[900:1500]$) and $N_T\in [17:22]$ at $L=24$
($N_J\in[500:1000]$).  For the PT cycles in $D$, $N_D\in[21:37]$,
parallel replicas at different $D$ were simulated, for size 
$L=6,8,10,12$ and $15$ ($N_J=1000$). In the latter case, to resolve
the coexistence region, varying
$\Delta D$ were used, larger in the pure phases and progressively
smaller approaching the transition.  The number of
Monte Carlo (MC) steps varies  from $2^{15}$ to $2^{21}$ according to $L$
and to the lowest
values of $T,D$ reached. \\
Thermalization has been cross checked by looking at: (i) the symmetry of the
overlap distributions $P_{N,J}(q)$, (ii) the $t$-log behavior of the
energy (when at least the last two points coincide), (iii) the lack of
variation of each considered observable (e.g., $P(q),\,P(\rho)$) on
logarithmic time-windows. \\
\begin{figure}[!t]
\begin{center}
\includegraphics[width=.5\textwidth]{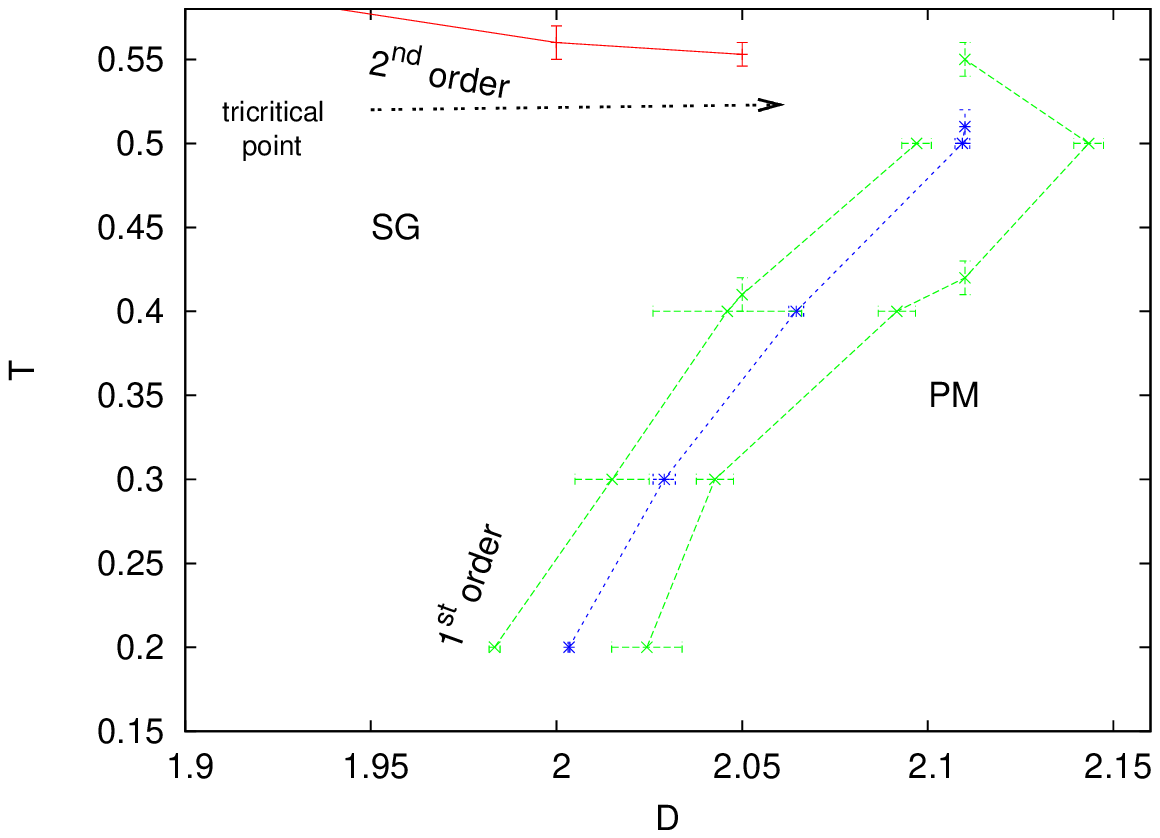}%
\includegraphics[width=.5\textwidth]{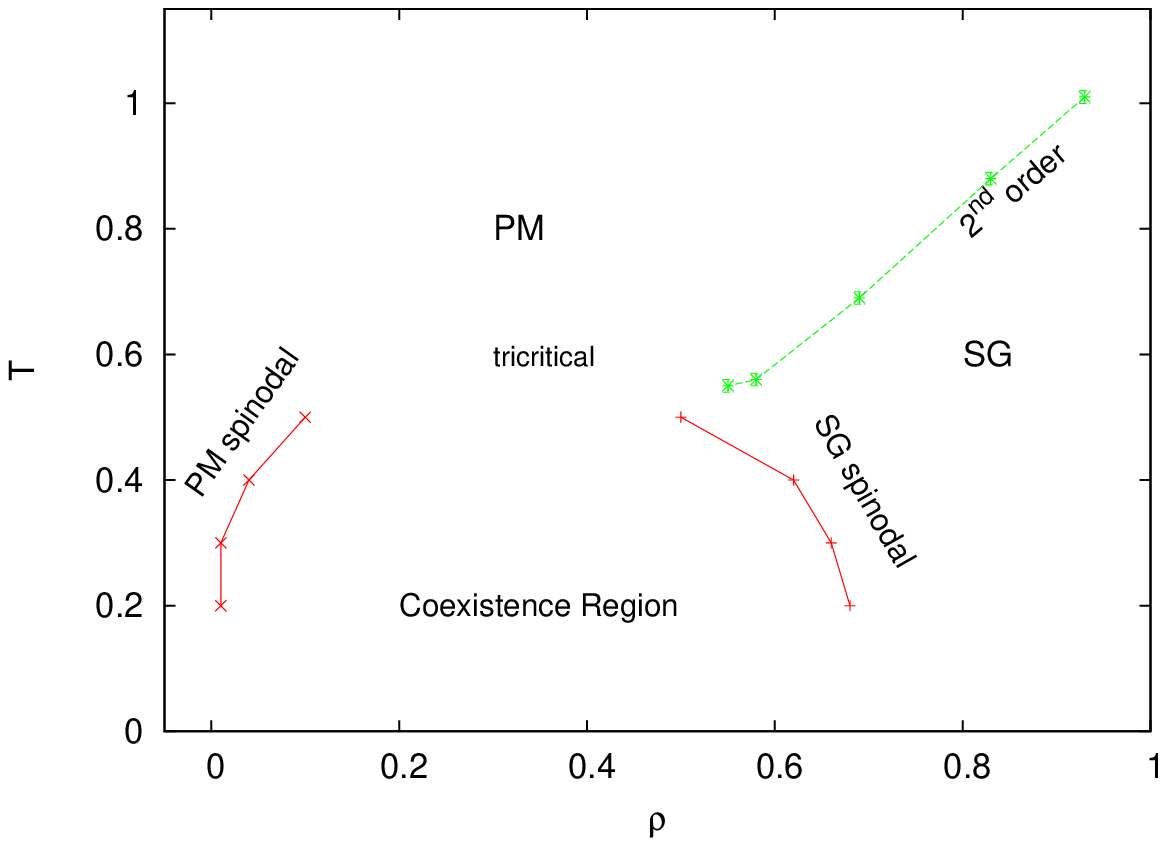}%
\caption{The phase diagrams of the BC model with
quenched disorder in three dimension. In
the left panel is shown the $T-D$ plane. 
The FOPT transition line develops 
a reentrance, it is an IT: decreasing
the temperature the system leaves the SG 
phase and liquefies in
a PM phase. 
In figure the spinodal lines (green) are reported
and the blue line is the critical line of the FOPT.
The arrow pointing to the region of
the phase diagram where is located the tricritical
point.
In the right panel the $T-\rho$ plane
is shown.%
}%
\label{phase}
\end{center}
\end{figure}
In fig. (\ref{phase}) the phase 
diagram of the model is shown:
both in $T-D$ and in $T-\rho$ plane. The
second order phase transition has
already been studied in \cite{Paoluzzi10}, 
details can be found in \cite{Leuzzi10}.
In short, by the study of the four-spins correlation
function 
\BEQ
C_4(x)\equiv\overline{\frac{1}{N}\langle \sum_i q_i^{(J)}q_{i+x}^{(J)}\rangle}=\overline{\frac{1}{N}\langle \sum_i s_i^{(1)} s_i^{(2)} s_{i+x}^{(1)} s_{i+x}^{(2)}\rangle}
\EEQ
we can introduce a correlation lenght $\xi(T,L)/L$
that is scale invariant at the critical point: this property 
allows to define a size-dependet $T_c(L)$. Finally, through
Finite Size Scaling techniques, it is possible to
calculate the critical temperature in the Thermodynamic
Limit\cite{palassini,BCPRB00}.

\begin{figure}[!t]
\begin{center}
\includegraphics[width=.35\textwidth,angle=270]{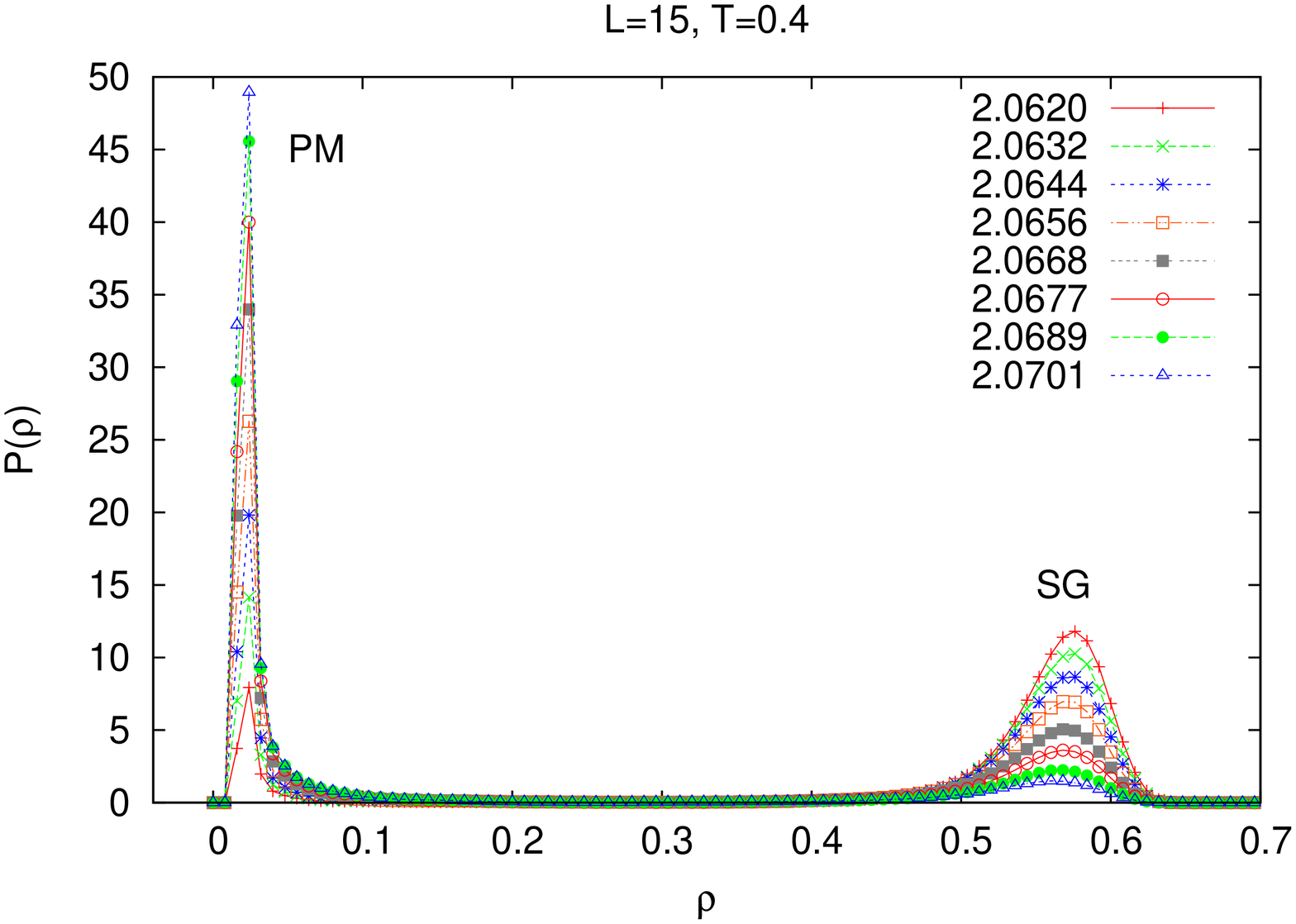}%
\includegraphics[width=.35\textwidth,angle=270]{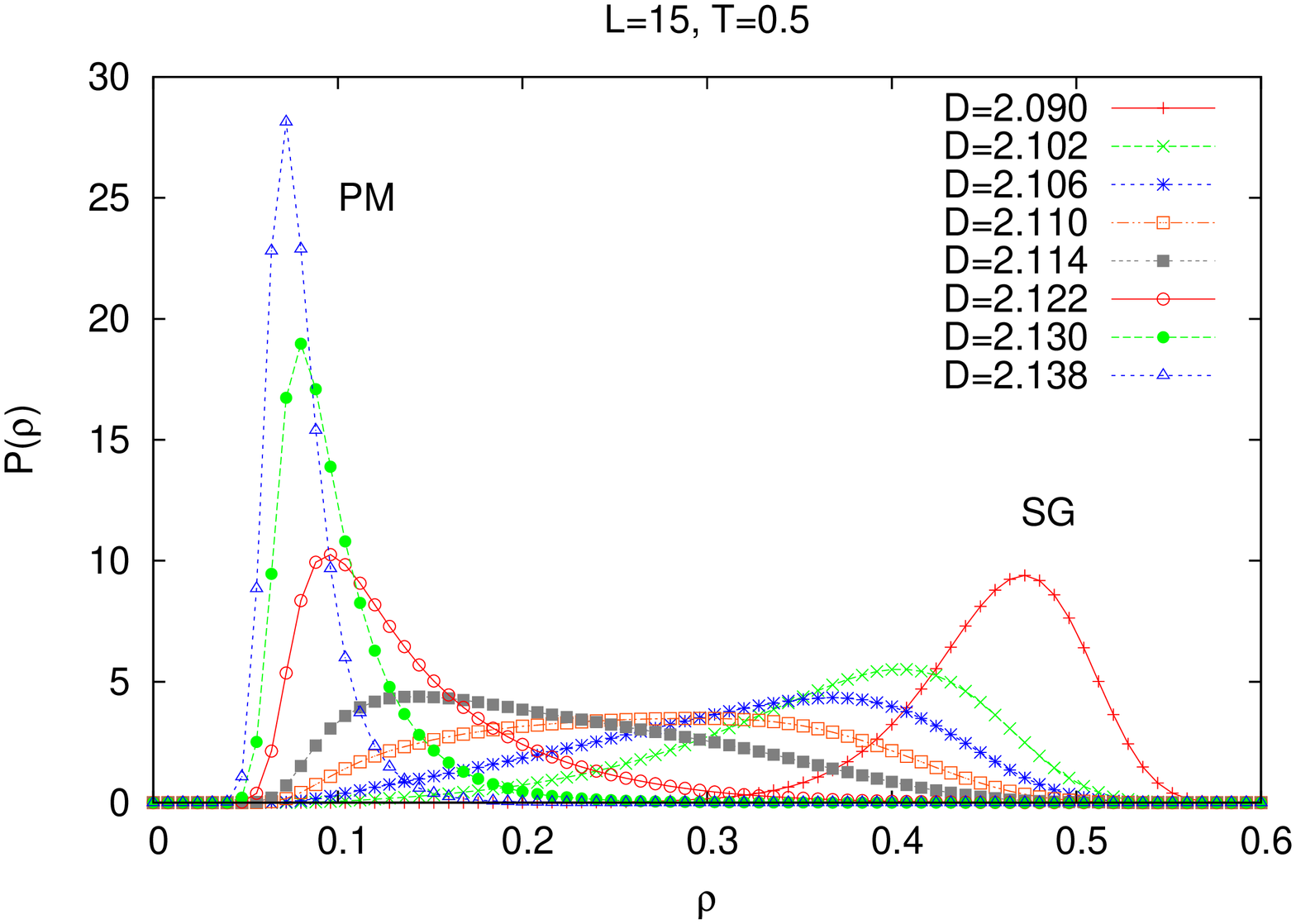}%
\caption{Density distribution $P_L(\rho)$, $L=15$, across the
coexistence region at $T=0.4$ (left) and $T=0.5$ (right): two peaks develop at $\rho_{PM}$ and
$\rho_{SG}$. As $D$ increases the thermodynamically relevant phase
(lowest free energy) passes from SG to PM in a first order phase
transition. The dominant phase correpsonds to the one with larger
probability, i.e., larger intergral of the peak. As the peak at
$\rho_{SG}$ vanishes the system is in a purely PM phase.
At $T=0.5$ (right) a coninuum part between the SG and PM peaks
appears: the distribution of the densities of two phases are
overlapping.  
}
\label{letter}
\end{center}
\end{figure}

The order parameter that drives the FOPT at finite $L$
is the density distribution $P(\rho)$: varying $D,T$, the system
undergoes a transition with a discontinuous jump in $\rho$ (and, thus,
in $q$). The system is in the coexistence region if $P(\rho)$ displays
two peaks corresponding to the PM (low $\rho$) and SG (high $\rho$)
phases.
The first order transition line $D_c(L,T)$, is the locus of points where the
two phases are equiprobable, i.e., the areas of the two peaks are
equal \cite{Hill}:
\BEQ 
\int_{0}^{\rho_0} d\rho\,P(\rho)=\int_{\rho_0}^1
 d\rho\, P(\rho) \EEQ 
where $\rho_0\in [\rho_{PM}:\rho_{SG}]$
such that $P(\rho)=0$ (or minimal next to the tricritical point).
%
 %
In order to determine the transition point a method is 
to compare the areas under
the distributions, cf. Eq (10). This is the point at which the
configurations belonging to the SG phase and those belonging to the PM
phase have the same statistical weight: they yield identical
contribution to the partition function of the single pure phase, and
the free energies of the two coexisting phases are equal.  We work at
finite T moving D in that region and this method works quite well for
$T\leq0.4$ because the two peaks are clearly separated as soon as they
appear, cf. fig. (\ref{letter}). Determination of $D_c$ is robust
against reasonable changes of $\rho_0$. At T=0.5 we
have the problem that the distributions of the densities of the
two phases are overlapping. In that case, seen the arbitrariness of
choosing $\rho_0$, we actually determine the transition point as the D
value at which the peaks have the same height.\\ 
%
%
To have a better confidence with the results, two other
methods have been used to determine
%
the first order transition. 
These are not plagued by the problem of dealing with
overlapping distributions since they do only rely on averages. The methods
are:
\begin{enumerate}
\item[(i)]  {\itshape Equal distance}:
at a given T, we plot D {\itshape versus} average
$\rho$, we extrapolate the $D(\rho)$ curves both from
the PM and the SG phase ($D_{PM}(\rho)$ and $D_{SG}(\rho)$) 
and we make a
Maxwell-like construction determining a value of $D_c$ at which
\BEQ
\rho(D_c) = \frac{1}{2}\left(\rho_{\rm PM}(D_c) + \rho_{\rm SG}(D_c)\right)\,.\EEQ  
\item[(ii)]  {\itshape Equal area}: 
equivalently (the equivalence is in the thermodynamic limit)
one can determine $D_c$ as the value at which 
\BEQ \int_{D_{\rm SG}}^{D_c} \rho_{\rm PM}(D) dD + \int_{D_c}^{D_{\rm
PM}} \rho_{\rm SG}(D) dD = \int_{D_{\rm SG}}^{D_{\rm PM}} \rho(D) dD
\EEQ 
where $D_{\rm SG}$ and $D_{\rm PM}$ are arbitrary, provided they
pertain to the relative pure phases. The extrapolated, $\rho_{\rm
PM}(D)$ ($\rho_{\rm SG}(D)$) is the inverse of the extrapolated curve $D_{\rm
PM}(\rho)$ ($D_{\rm SG}(\rho)$). 
The curves are obtained by using all data,
including those in the candidate coexistence region.

We will show and compare in Sec. (3.2) the results obtained by these three methods.
\end{enumerate}
\subsection{Second Order Phase Transition}
\begin{figure}
\begin{center}
\includegraphics[width=.35\textwidth,angle=270]{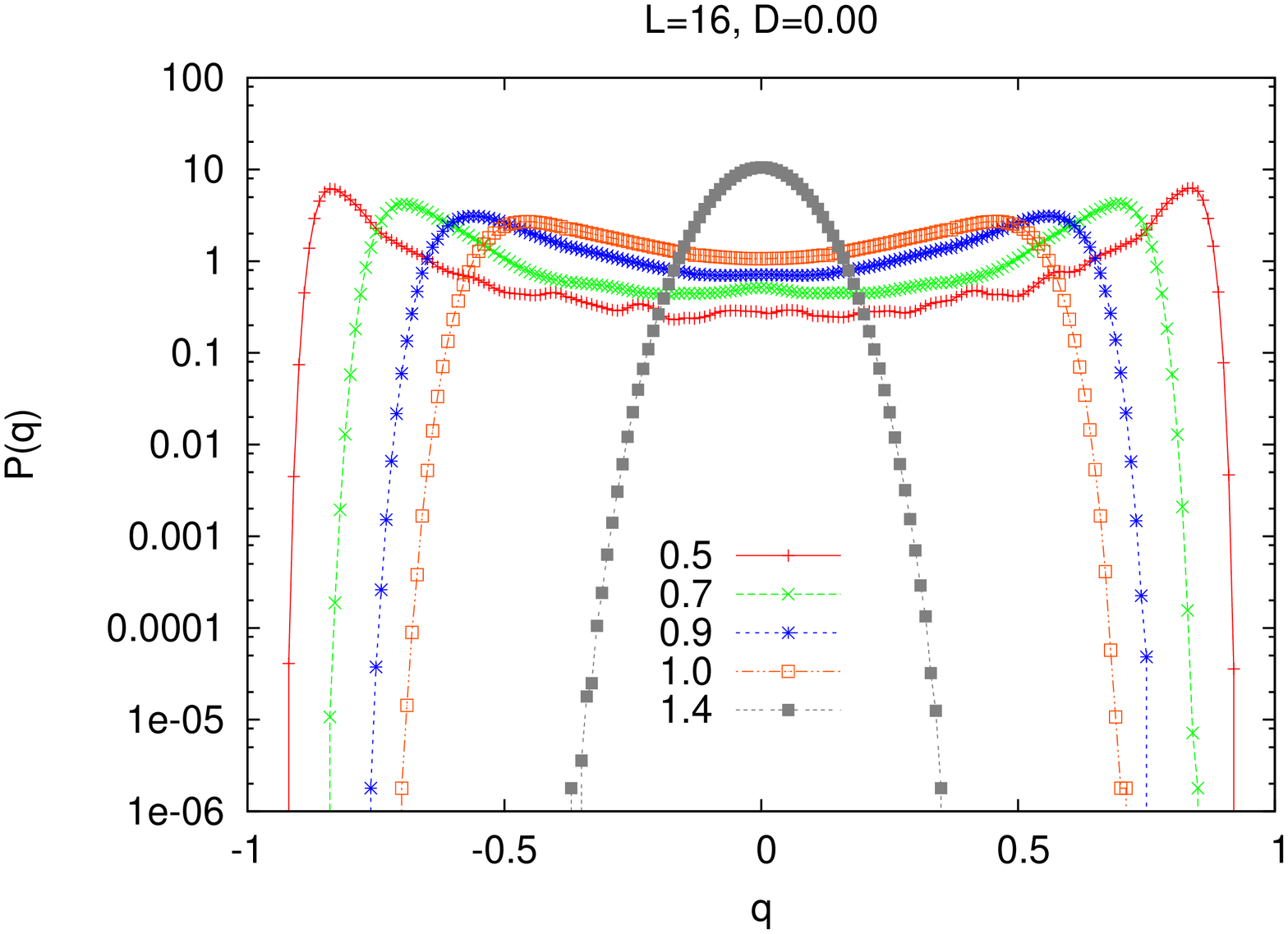}%
\includegraphics[width=.35\textwidth,angle=270]{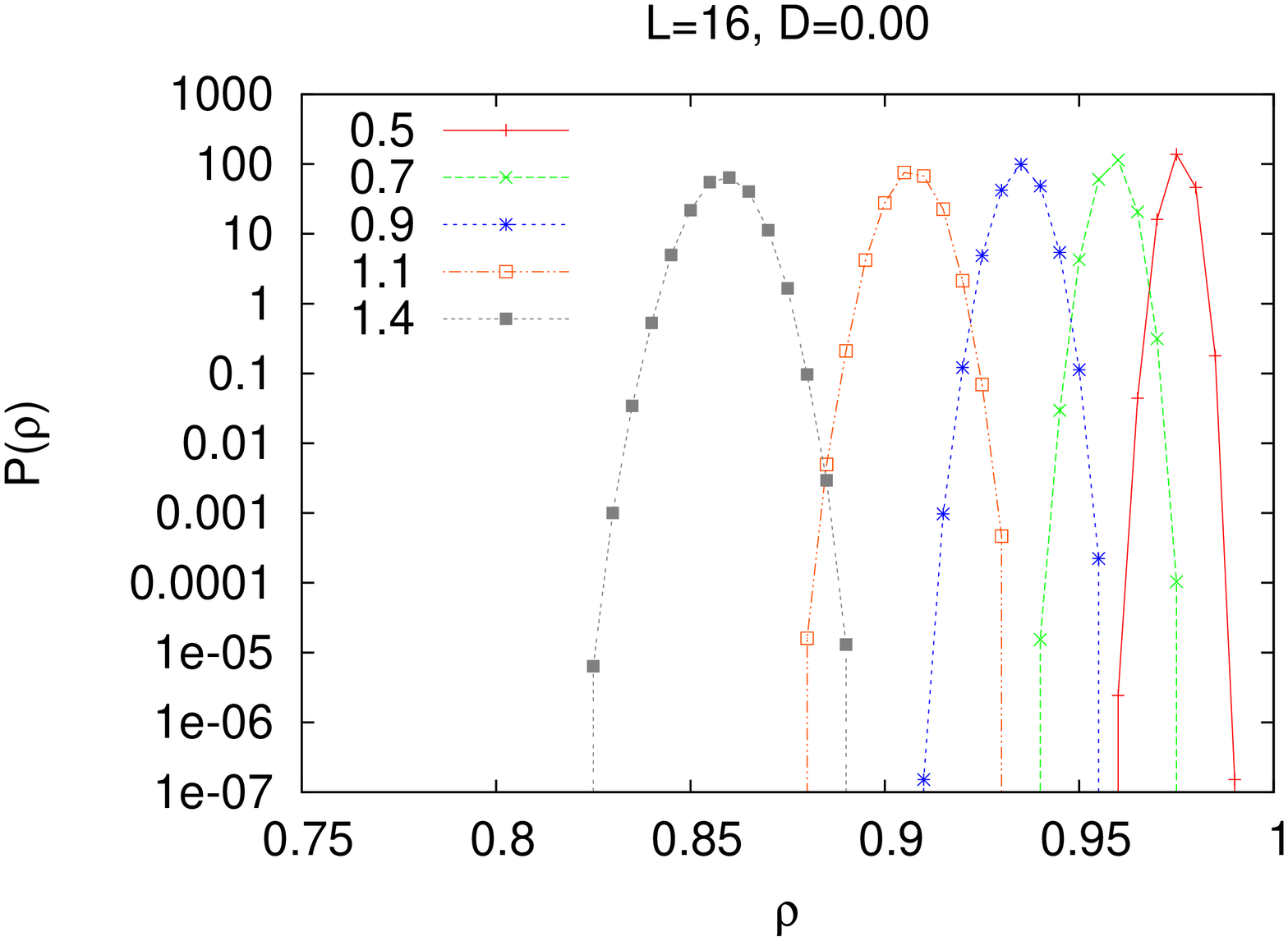}%
\caption{%
$P_L(q)$ and $P_L(\rho)$ for $L=16$ and $N_J=1500$ at $D=0.00$.
Across the transition $P(q)$ (left
panel) changes shape continuously from a Gaussian 
(in the high temperature phase) to, decreasing 
the temperature, a double peaked distribution
with a continuum part between the peaks. 
In this region of the
phase diagram the FOPT does not take place: $P(\rho)$
(right panel) does not change shape.
}%
\label{SOPT_II}
\end{center}
\end{figure}

In fig. (\ref{SOPT_II}) (left panel) we report
the distribution of overlap $P(q)$
for a system of linear size $L=16$ and $D=0$: it changes 
shape accross the second order phase transition\cite{MPR98} between 
the PM and SG phase from
a Gaussian to a double 
peaked distribution 
\BEQ 
P_{SG}(q)\propto \delta(q-q_{EA})+
\delta(q+q_{EA})+f(q,L)\; ,
\EEQ
where $f(q,L)$
is a continuous function depending on the size. 
%
In the right panel of the figure we
show the behavior of $P(\rho)$ 
at fixed values of the crystal field
and several temperatures: decreasing
the temperature, deep in the SG phase,
the average number of active sites is close
to one
\BEQ
\lim_{T\to 0} \overline{ \langle \rho \rangle} = \int_0^1 \,d\rho \, P(\rho,T) \, \rho \sim 1\, .
\EEQ 
The distribution $P_L(\rho)$
does not change shape across this transition.
\subsection{First Order Phase Transition}
Beyond the tricritical point, FOPT takes place and 
the system undergoes a discontinuous transition between
an ``inactive'' PM phase ($\overline{\langle\rho\rangle}\equiv \rho_m\sim 0$)
and an ``active'' SG phase ($\rho_m\neq 0$). 
In the coexistence region, we can write the $P(q)$
as sum of two contributes:
\BEQ
P(q)=P_{SG}(q)+P_{PM}(q)\, .
\EEQ
For the PM contribution $P_{PM}(q)$ 
we have a Gaussian strongly peaked around
$q=0$.
The $P_{SG}(q)$ consists of a double peak (trivial)
distribution with a continuum (non trivial) part
between the two peaks, cf. Sec. 3.1.
%
\begin{figure}[!t]
\begin{center}
\includegraphics[width=.35\textwidth,angle=270]{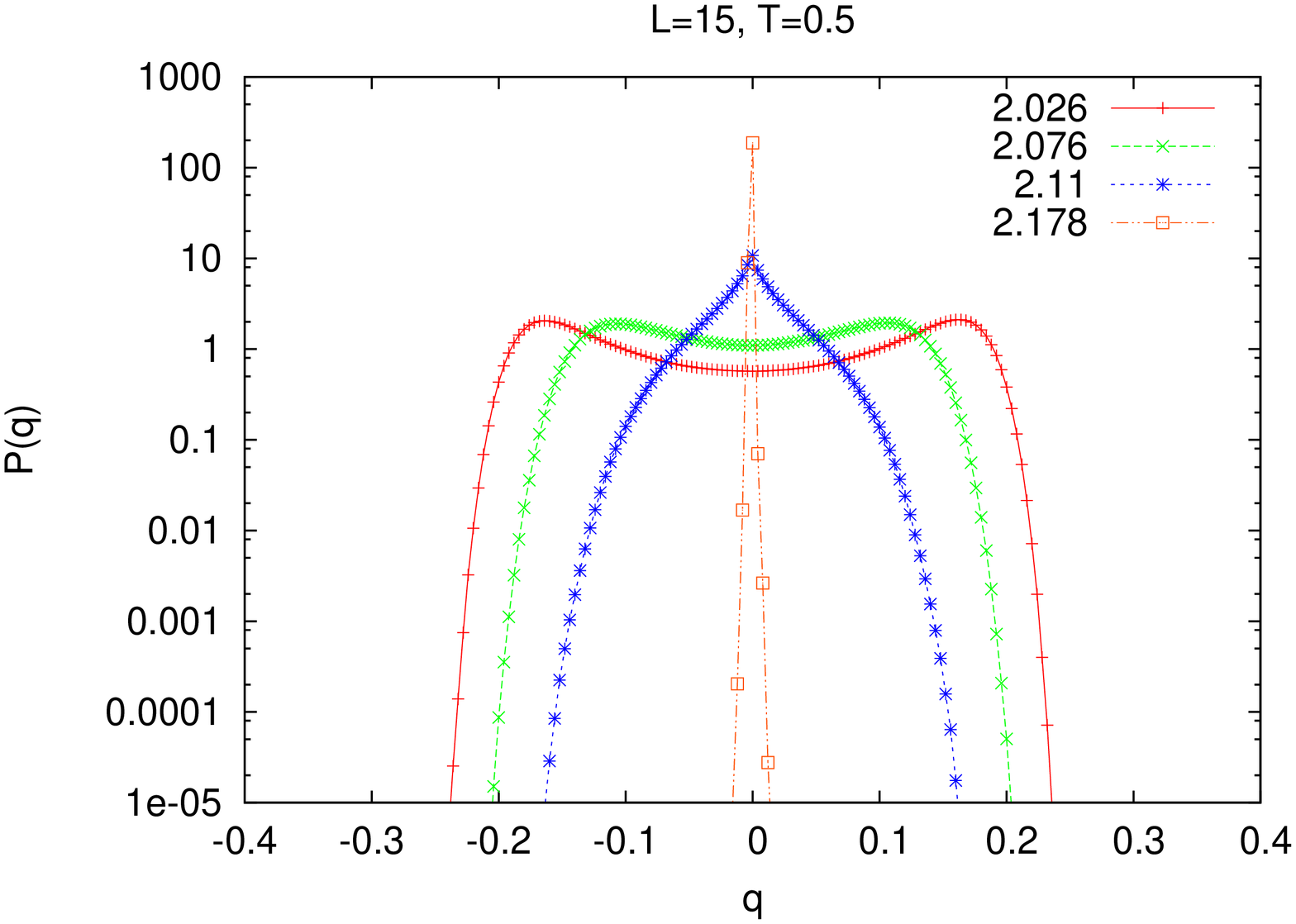}%
\includegraphics[width=.35\textwidth,angle=270]{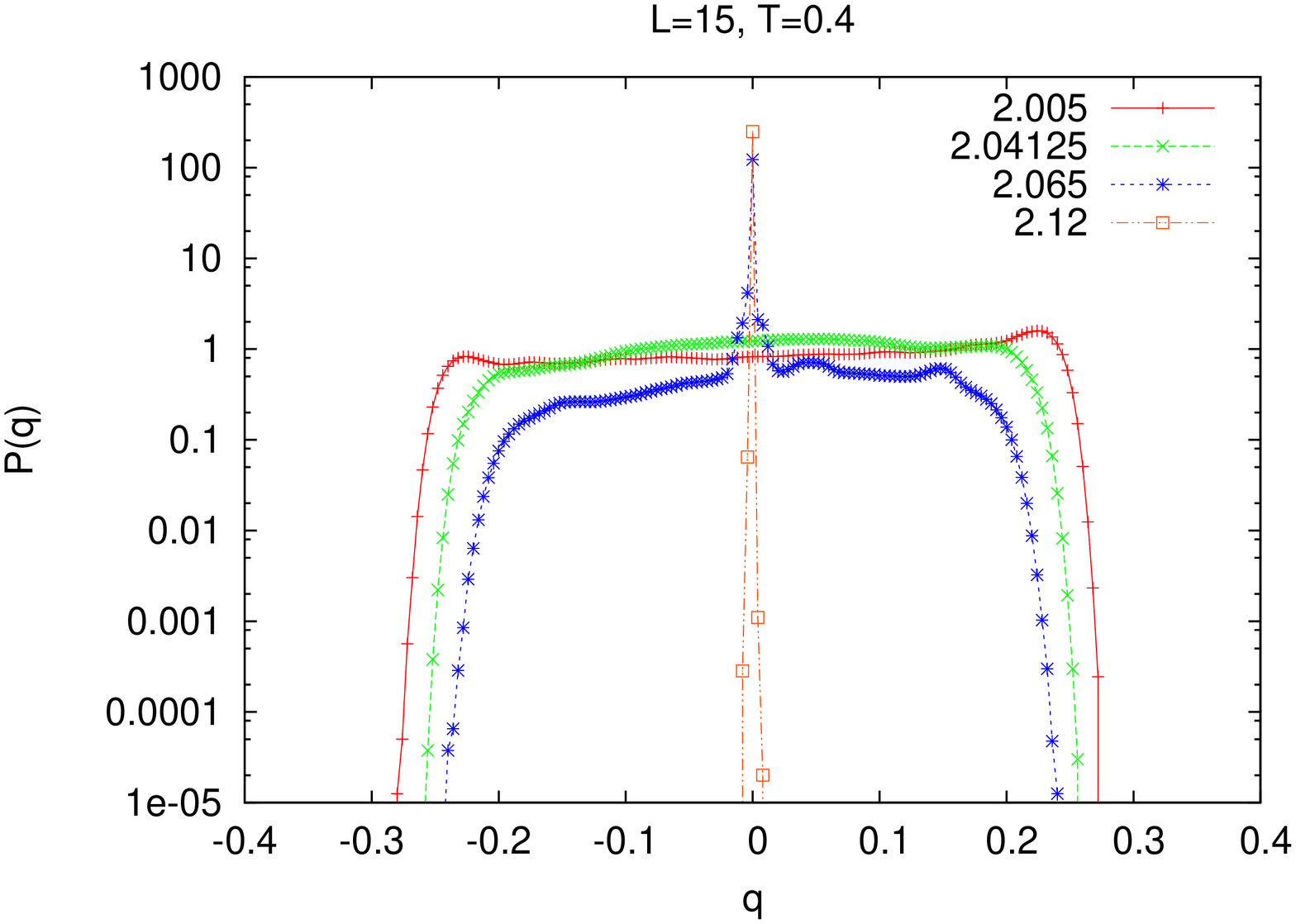}\\%
\includegraphics[width=.35\textwidth,angle=270]{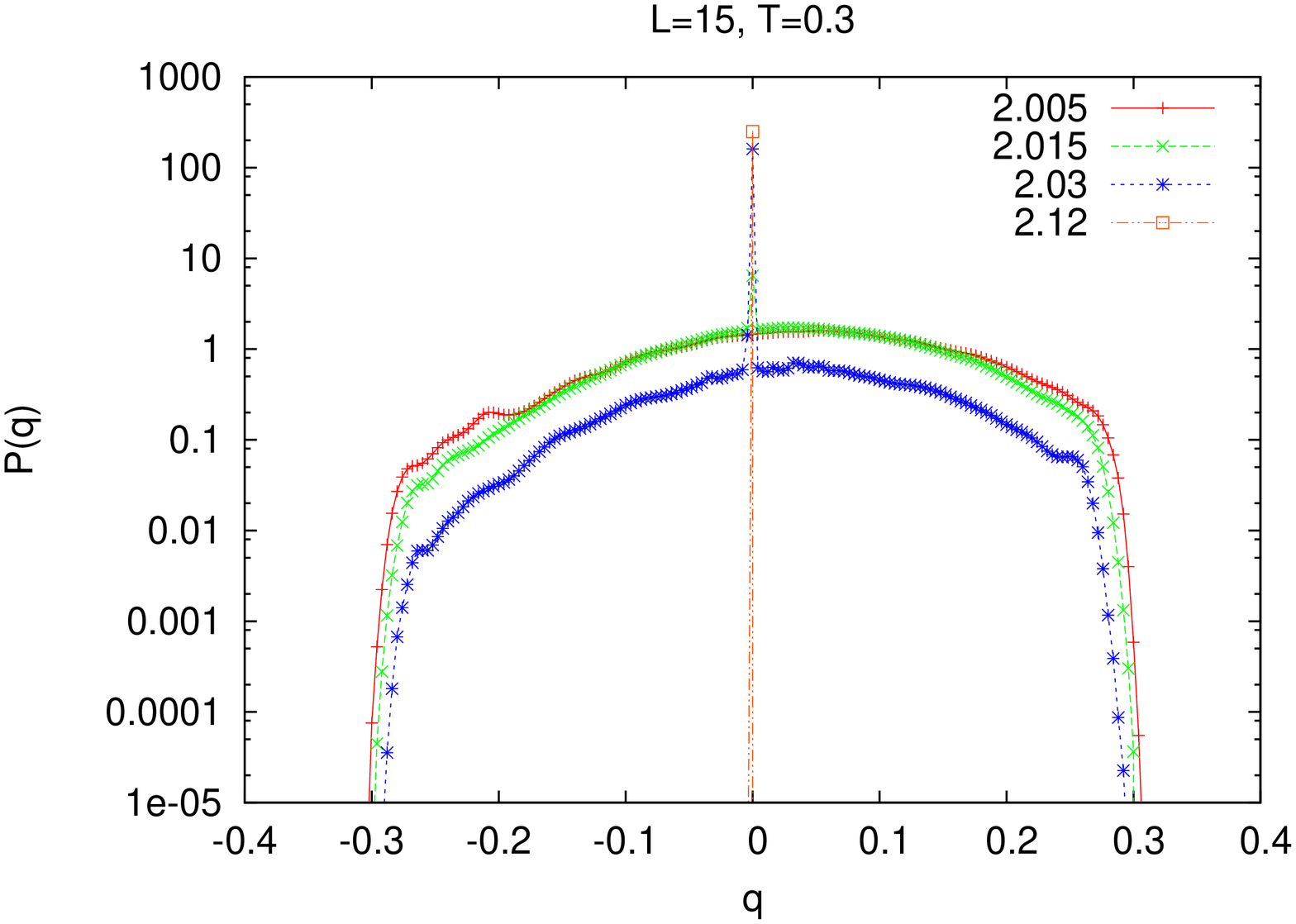}%
\includegraphics[width=.35\textwidth,angle=270]{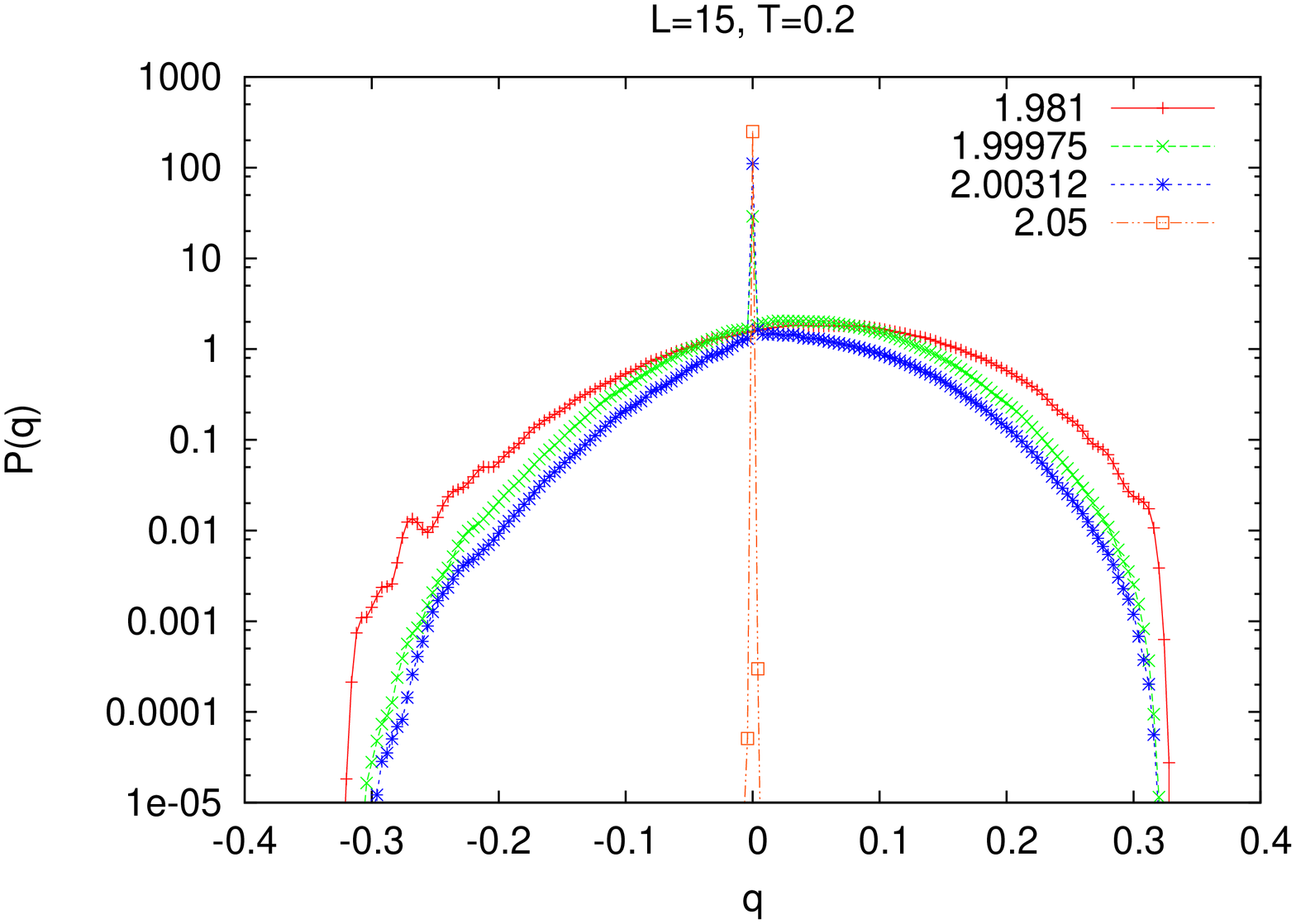}%
\caption{ $P(q)$ across the FOPT at fixed $T$
and different $D$ values: we have
a coexistence region between the phases
SG and PM. The PM phase contributes
to the $P(q)$ distribution with a delta
function in $q=0$.%
}%
\label{FOPT_q}
\end{center}
\end{figure}
\begin{figure}[!t]
\begin{center}
\includegraphics[width=.35\textwidth,angle=270]{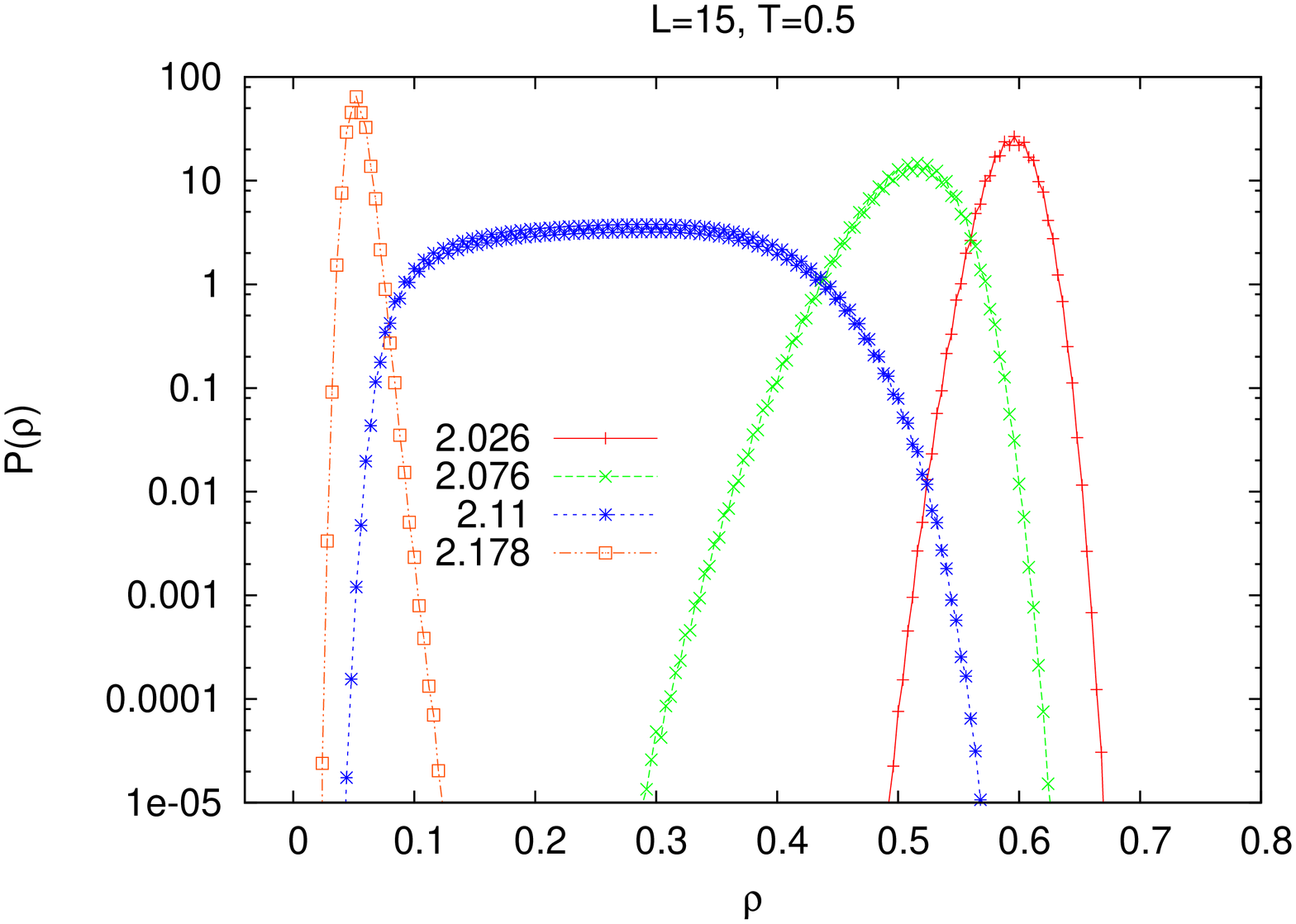}%
\includegraphics[width=.35\textwidth,angle=270]{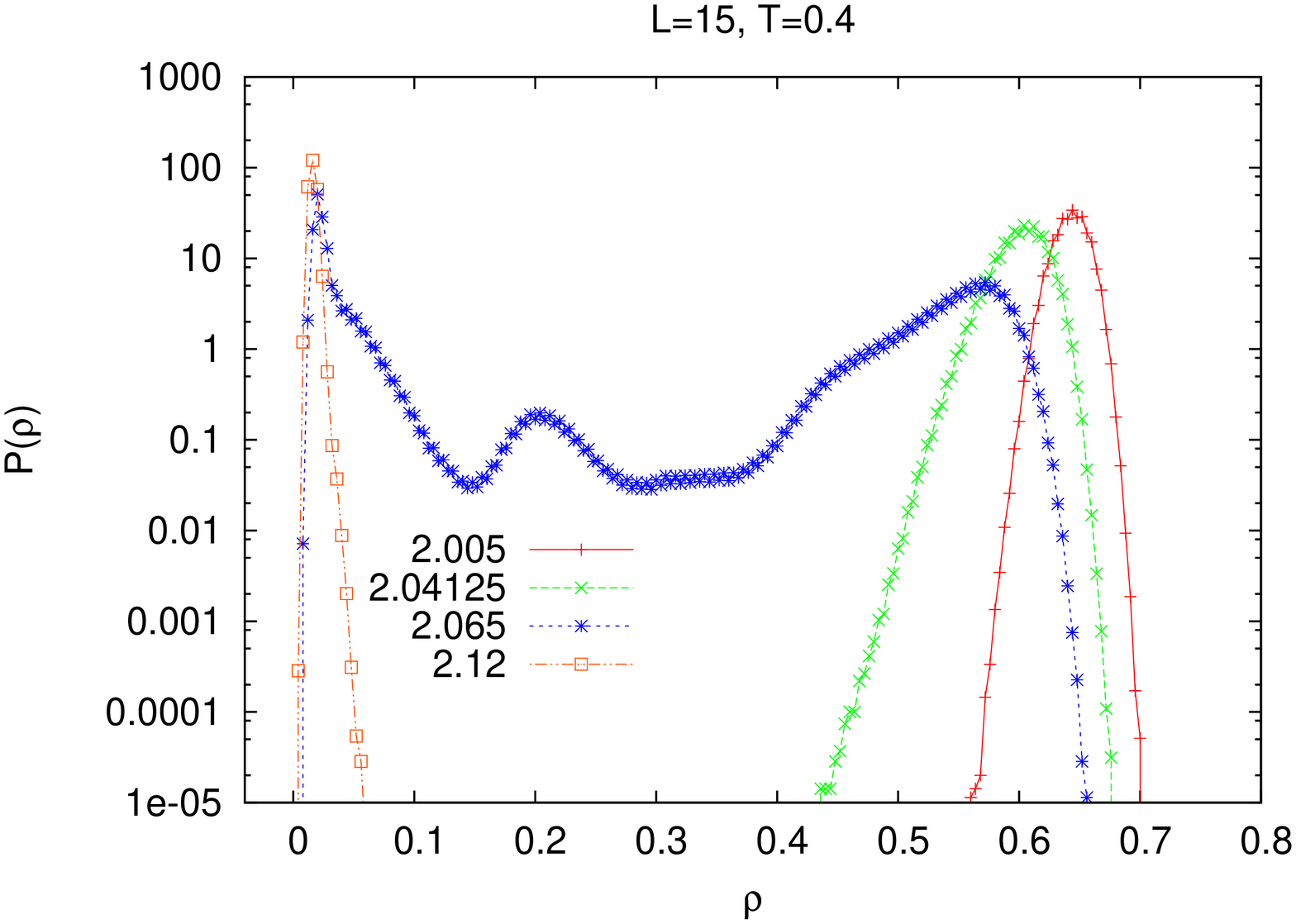}\\%
\includegraphics[width=.35\textwidth,angle=270]{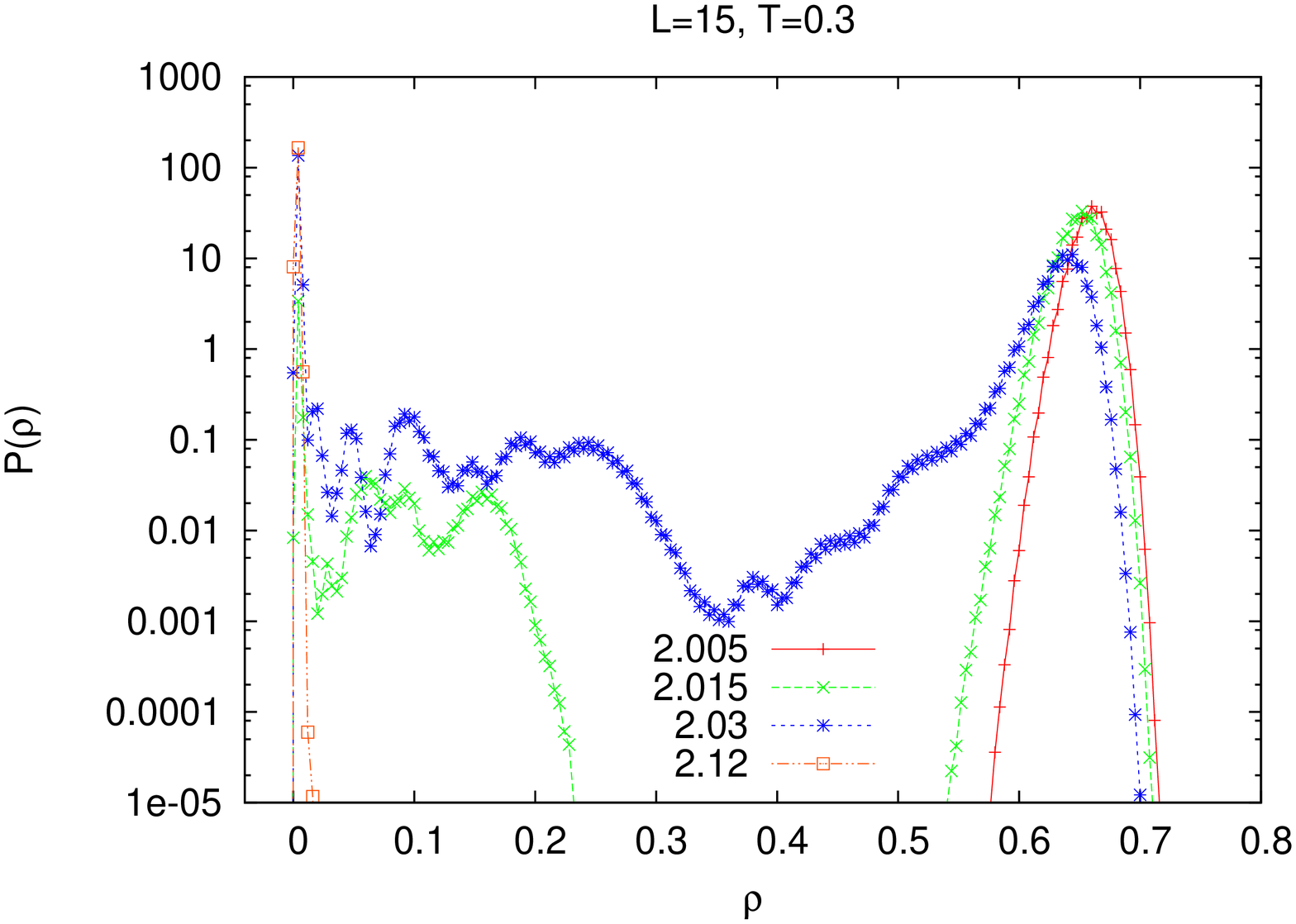}%
\includegraphics[width=.35\textwidth,angle=270]{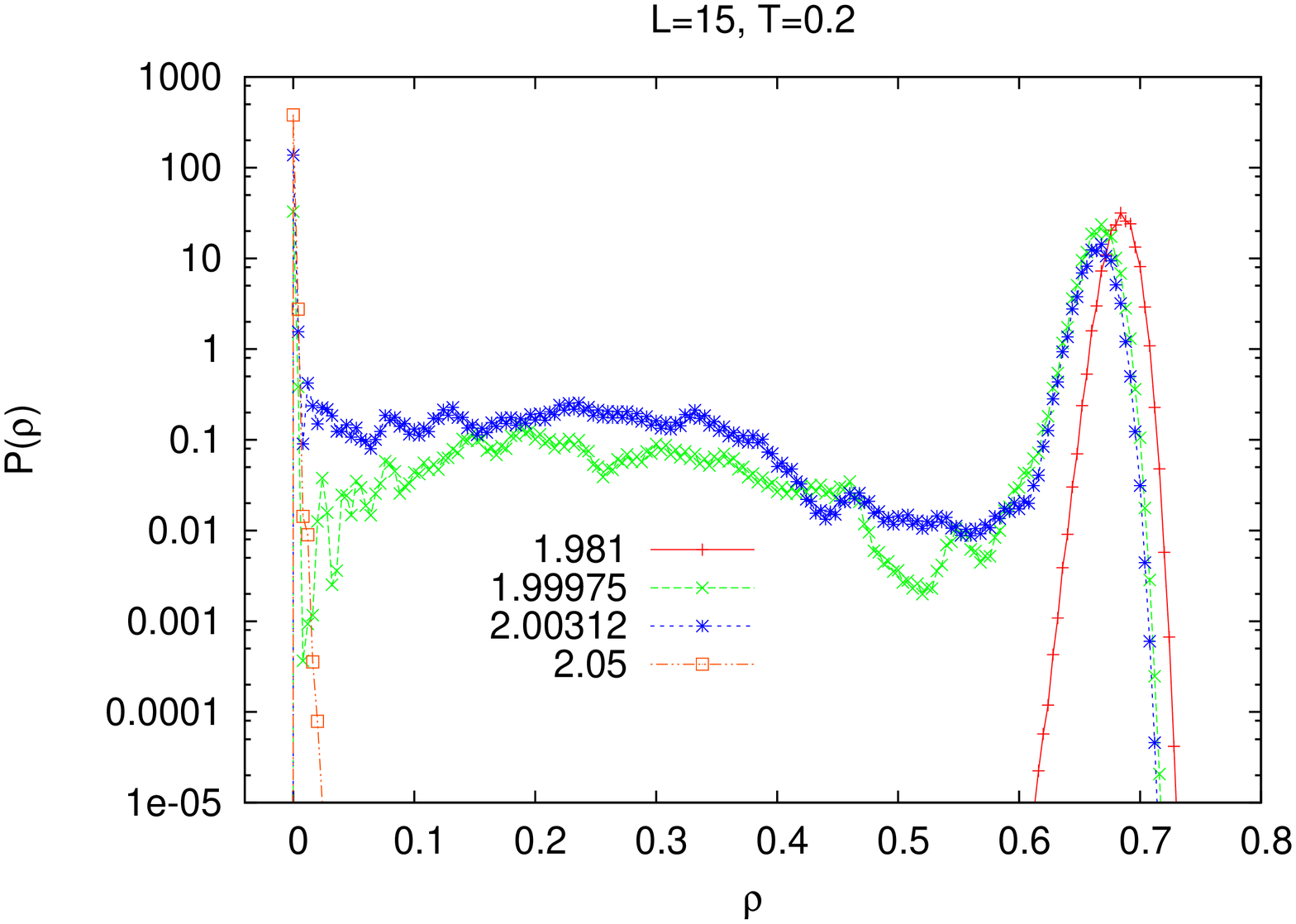}%
\caption{ $P(\rho)$ during the FOPT: a 
double peak signal the
coexistence of phases while,
when the system is in a pure phase,
we have a single peak.%
}%
\label{FOPT_rho}
\end{center}
\end{figure}
In fig. (\ref{FOPT_q}) we show the  behavior
of $P(q)$ at different temperatures when the FOPT
occurs. In the coexistence region, besides
the double peak with a continuous part of the
SG phase, a peak in $q=0$
appears due to the large density of
empty sites. 
\\
As shown in fig. (\ref{phase}), cf. also
Sec. 2, FOPT occurs as an
IF: 
the PM phase riches of inactive sites, below the T-range at which a
pure SG phase is present,
becomes the low
temperature (and less entropic) phase.
This can be better observed in fig. (\ref{FOPT_rho})
where $P(\rho)$ is represented at several values of $D$ through the transition.
In table (\ref{tab}) the critical values of the FOPT reported
are obtained by the equal weight, equal area and equal distance methods.
\begin{table}
\centering
\begin{tabular}{| l | c | c | c | c | c |}\hline 
T  	&	$D_c[P(\rho)]$  &$D_c[\rho_m]$	&$D_c[\mathcal{A}(\rho)]$&	$D_{sp}(PM)$ 	&	$D_{sp}(SG)$	\\ \hline
0.2    &	1.992(2) 		&1.998(3)			&1.999(2)				&	1.98333(15) 	& 	2.0243(95) \\   
0.3   	&  	2.032(2) 	  	&2.032(3)			&2.030(1)				&	2.015(1)   		&  	2.043(5) 	\\
0.4   	&  	2.061(1) 	  	&2.060(2)			&2.058(1)				&	2.046(2)   		&   	2.092(5) 	\\
0.5   	&  	2.107(1) 	  	&2.102(1)			&2.102(2)				&	2.097(4)  		&  	2.143(4) 	\\ \hline
\end{tabular}
\vspace{0.2cm}\caption{Critical values of the field $D$ calculated through the equiprobability
of the phases (eq. 9, blue line in the left panel of fig. (\ref{phase})), the equal distance (eq. 10) and the equal area (eq. 11) respectively. 
In the last two columns are
reported the spinodal lines $D_{sp}$ (green lines in the left panel of fog. (\ref{phase})). 
}
\label{tab}
\end{table}

From the study of the shape of $P(\rho)$
in the different phases (high-T PM phase,
SG phase and low-T PM phase), the 3D BC-random
model, allows us to interpretate the IF scenario
in a very intuitive
picture.
Starting from a point of the phase diagram (D,T)
in the high temperature phase (high-T and D$>0$)
the PM phase, as it can be seen in fig. \ref{SOPT_II}, is dominated
by the active sites with $s_i^2\neq 0$ while
$\overline{\langle q_i^{(J)} \rangle}=\langle s_i^{(1)} s_i^{(2)} \rangle=0$.
Decreasing the temperature a Second Order Phase
Transition takes place: $P(\rho)$ does not
change shape and $\rho_m(PM)=\rho_m(SG)$.
Decreasing further the temperature, the system
undergoes to an Inverse-FOPT: $P(\rho)$ becames
a double-picked distribution. The coexistence
of the phases occurs between a SG ($\rho_m>0$) phase
and a PM ($\rho_m\sim 0$) phase. Since the
configuration space is dominated by 
inactives sites, the entropy of
the PM phase becames smaller than 
the entropy of SG phase, leading to
IF.

\section{Conclusions}
To conlcude, we analyzed the phenomenology of the
BC-random in three dimensions 
studying the behavior of the order
parameter distributions: $P(q)$ and
$P(\rho)$. In the case of the continuous
transition, the lattice has an high 
density of active sites already the 
PM phase:\footnote{In the limit $\lim_{T\to 0, D\to - \infty} \int_0^1 \,d\rho P(\rho,T,D)\rho=1$
and we obtain the Edward-Anderson model\cite{ea}.}
\BEQ
\lim_{T\to 0, D=0} \int_0^1 \,d\rho P(\rho,T,D)\rho\sim1
\EEQ
and the density changes with continuity across the transition,
whereas $P(q)$ changes shape from a Gaussian to a double
peaked distribution.
When the discontinuous First Order Phase Transition takes place,
$P(\rho)$ becomes a double peaked distribution due
to the coexistence of $PM$ and $SG$ phases. The transition
is an IF and the order parameter $\rho$ jumps discontinuously
between a low density and poor interacting PM phase (less
entropic\cite{Paoluzzi10}) to an high density SG phase.
Through the study of $P(q)$ we have a clear
evidence of the coexistence of a PM phase, riches in empty sites
($\rho_m\to 0$) where the distribution has a single
peak, and a SG phase with $\rho_m \neq 0$ with
a double peaked $P(q)$ and a continuum part between the
peaks. 

Finally we observe that the BC-random in three dimension is one
of the few short-range disorder system
undergoing a thermodynamic FOPT. 
We have found in literature
only one system that displays a FOPT:
the $4-$Potts glass studied by Fernandez
{\itshape et al.}
\cite{Fernandez08}.  In that case,
though, randomness tends to strongly smoothen the transition into a
second order one and the finite size effects are very
strong in determing the tricritical point.
Our study, thus, confirms in a clear way
the claim
of the existence of a FOPT
in presence of quenched disorder 
thanks to almost negligible finite
size effects.\\ 
Moreover, we notice that the FOPT of the BC-random
is driven only by external thermodynamics
parameters, e.g., the temperature and the
chemical potential.
Even though, from the point of view of the numerical
simulation changing the pressure, bond
dilution\cite{Toldin09} or even the relative probabilities
of having ferro- or antiferro-magnetic interactions
\cite{Fernandez08} is technically equivalent, the latter are
complicated to control in a real experiment
and require the preparation of several
samples with different microscopic properties. 
Eventually, we are not privy to any three dimension
short-range system with quenched disorder
undergoing an IF.
%


\end{document}